\documentstyle[12pt]{article}
\title{\bf
	     STOCHASTIC RESONANCE IN  3D ISING FERROMAGNETS}
\author{
{\bf   Zolt\'an N\'eda}\thanks
{Permanent address: {\it
      Babe\c {s}-Bolyai University , Dept. of  Physics,
      str. Kog\u {a}lniceanu 1, RO-3400 Cluj-Napoca, Romania }  } \\
{\small\it
       University of Bergen, Section for Theoretical Physics }\\
{\small\it
	 All\'egaten 55, N-5007 Bergen, Norway}  }

\date{}

\begin{document}

\maketitle

\begin{center}

Abstract\\
\end{center}

Finite 3D Ising ferromagnets are studied in periodic magnetic fields both by
computer simulations and mean-field theoretical approaches. The phenomenon of
stochastic resonance is revealed. The characteristic peak obtained
for the correlation function
between the external oscillating magnetic field and magnetization versus
the temperature of the system, is studied for various external fields and
lattice
sizes. Excellent agreement between simulation and theoretical results are
obtained.

\vspace{0.5in}
PACS number(s): 05.40.+j; 75.10.-b; 02.70.Lq

\vfill \eject

\section{Introduction}

It is well-known [1-4], that periodically modulated bistable systems in
the presence of noise exhibits the phenomenon of stochastic resonance
(SR). The bases of this phenomenon is that the correlation $\sigma$ between
the modulation signal and the response of the system presents an extremum for
a given noise intensity.

Evidences for SR were found in analog simulations with proper electronic
circuits [3,5,6], laser systems operating in multistable conditions [7],
electron paramagnetic resonance [8,9], in a free standing magnetoelastic
ribbon [10] or in globally coupled two-state systems [11,12]. Theoretical
aspects of the problem were reviewed in [4].

Considering a special and practically important case of coupled two-state
systems, recently we reported on the possibility of obtaining the SR in
finite two dimensional Ising systems [13]. In the mentioned work, in contrast
with
all other earlier works we did not consider any external stochastic forces,
only
the thermal fluctuations in the system. The problem was studied by computer
simulations considering a heat-bath dynamics.

The present paper intends to complete the earlier one [13], studying the
important
three dimensional (3D) case. Presuming that in 3D the mean-field theories give
reasonable results, we also consider a theoretical approach to the phenomenon.
We discuss and compare our theoretical and computer simulation results.

\section{The Method}

For detecting SR we need a system in a double well potential, governed by a
stochastic force. In the meantime the two minima of the double well potential
must be modulated periodically in antiphase. One can immediately recognize that
ferromagnetic Ising systems in oscillating magnetic fields satisfy all these
conditions:
\begin{itemize}
\item at zero thermodynamic temperature the free-energy versus magnetization
curve
has the double well form,
\item an external periodic magnetic field [$B(t)$], modulates the two minima in
antiphase,
\item the effect of a positive temperature can be viewed as a stochastic
driving force,
\item the magnetization as a function of time ($M(t)=\sum_i S_i^z$) can be
considered as the response function of the system.
\end{itemize}
Due to the fact that in the proposed system the noise intensity (thermal
fluctuations)
is temperature dependent, the characteristic maximum for SR must be detected at
a
resonance temperature ($T_r$) in the plot of the correlation $\sigma$
($\sigma=\mid<B(t) \cdot M(t)> \mid$) versus the temperature (T).

The Hamiltonian of our problem is
\begin{equation}
H=-J \sum_{i,j} S_i^z S_j^z + \mu_B B(t) \sum_{i} S_i^z,
\end{equation}
where the sum is referring to all nearest neighbours, $S_i^z=\pm 1$, $\mu_B$ is
the
Bohr magneton and $B(t)$ the external magnetic field. We consider $B(t)$ in a
harmonic form:
\begin{equation}
B(t)=A \sin{(\frac{2 \pi}{P} t)}.
\end{equation}

We mention that recently 3D Ising systems in oscillating magnetic fields were
already
considered by computer simulations [14]. However in [14] the authors study  the
hysteretic response of the system and no evidences for SR are discussed.

We will study the proposed system (1) from the viewpoint of SR both by computer
simulations and by a theoretical approach.

\subsection{Computer Simulation}

To study the time evolution of the proposed system (1) first we considered  a
computer simulation with heath-bath dynamics. The time scale was chosen in a
convenient way, setting the unit-time interval equal with the average
characteristic
time ($\tau$) necessary for the flip of a spin. We have taken this time
interval $\tau$ as
constant, and thus independent of the temperature. Although this assumption is
just a working hypothesis we expect to give useful qualitative results. The
spin flips
were realized with the probabilities of the Metropolis [15] algorithm, choosing
the spins
randomly at each moment.

The simulations were performed on cubic lattices with $W=N\times N \times N$
spins,
considering the value of $N$ up to $50$. One simulation step was defined as $W$
trials
of changing spin orientations, and corresponds to a time interval $\tau$. The
period
($P$) of the oscillating magnetic field will be also given in these $\tau$
units.
The amplitude $A$ of the magnetic field is considered already multiplied by
$\mu_B/k$, and thus will have the dimensionality of the temperature ($k$ is the
Boltzmann constant). The temperature will be considered in arbitrary units, and
the critical temperature of the infinite system ($T_c\approx 4.444 \:J/k$) will
be
set to $100$ units. Starting the system from a random configuration we
considered
$10 000$ simulation steps to approach the dynamic equilibrium. The correlation
function between the driving field $B(t)$ and the magnetic response M(t)
\begin{equation}
\sigma=\mid <B(t) \cdot M(t)> \mid= \mid \frac{1}{n} \sum_{i=1}^{n} B(t_i)
M(t_i) \mid
\end{equation}
was studied after this during $10 000$ extra iterations. (The averaging in (3)
is
as a function of time, and $t_i=\tau \cdot i$).

The correlation ($\sigma$) was studied as a function of:
\begin{itemize}
\item the temperature (T),
\item the lattice size (N),
\item the amplitude of the magnetic field (A), and
\item the period of the oscillating magnetic field (P).
\end{itemize}

\subsection{Theoretical approach}

We propose now a mean-field like theory to describe the time evolution
of the system.

Let us consider that $P(M,t)\: dM$ is the probability for having the
magnetization of the system between the values $M$ and $M+dM$ at a time
moment $t$. We know from the classical fluctuation theories that for a system
in
thermal equilibrium with a heat-bath the probability $P(x)$ of getting into a
state $x$ is proportional with the
\begin{equation}
P(x)\propto e^{-\beta F(x)}
\end{equation}
factor, where $\beta=1/(kT)$, and $F(x)$ is the free-energy in state $x$.
Presuming that for a very short time interval ($dt$) and fixed $M_o$ the
changes in $P(M_o,t)$ are influenced mainly by the neighbouring values
$P(M_o+dM_o,t)$ and $P(M_o-dM_o,t)$, the time evolution of $P(M,t)$ can
be approximated as
\begin{eqnarray}
\frac{\partial P(M,t)}{\partial t}= - \frac{(e^{-\beta F(M+dM,t)}+e^{-\beta
F(M-dM,t)})}
{S} P(M,t)+ \\ \nonumber
+ \frac{e^{-\beta F(M,t)}}{S} [P(M-dM,t)+P(M+dM,t)],
\end{eqnarray}
where:
\begin{equation}
S=\sum_{\{M\}} e^{-\beta F(M,t)}.
\end{equation}
Up to second order terms one can consider:
\begin{eqnarray}
P(M+dM,t)=P(M,t)+\frac{\partial P(M,t)}{\partial M} dM+ \frac{1}{2}
\frac{\partial^2
P(M,t)}{\partial M^2} dM^2, \\  \nonumber
P(M-dM,t)=P(M,t)-\frac{\partial P(M,t)}{\partial M} dM+ \frac{1}{2}
\frac{\partial^2
P(M,t)}{\partial M^2} dM^2, \\  \nonumber
F(M+dM,t)=F(M,t)+\frac{\partial F(M,t)}{\partial M} dM+ \frac{1}{2}
\frac{\partial^2
F(M,t)}{\partial M^2} dM^2, \\  \nonumber
F(M-dM,t)=F(M,t)-\frac{\partial F(M,t)}{\partial M} dM+ \frac{1}{2}
\frac{\partial^2
F(M,t)}{\partial M^2} dM^2.
\end{eqnarray}
In this manner equation (5) will be written as
\begin{equation}
\frac{\partial P(M,t)}{\partial t}=C \{ \frac{\partial^2
P(M,t)}{\partial M^2} + [ \beta  \frac{\partial^2
F(M,t)}{\partial M^2}-\beta^2 (\frac{\partial F(M,t)}{\partial M})^2] P(M,t)
\},
\end{equation}
with:
\begin{equation}
C=\frac{e^{-\beta F(M,t)}}{S} dM^2 .
\end{equation}

A few remarks on the obtained approximative "master" equation (8) must be made:
\begin{itemize}
\item The equation is  reasonable from the point of view that for
$\partial F(M,t)/\partial t=0$, $P(M,t)=Const. \exp{[-\beta F(M,t)]}$ is a
stationary solution.
\item The equation will not necessary keep the normalization of $P(M,t)$. For
assuring that $P(M,t)$ will have the meaning of probability, one must
apply equation (8) together with the
\begin{equation}
\int_{-W}^{W} P(M,t) dM=1
\end{equation}
normalization condition at each time step. ($W$ is the number of spins in the
system.)
\item The value of $C$ determines the speed of changing $P(M,t)$ at each
moment.
In reality $C$ is temperature dependent, but as a first approximation one can
consider
it constant. Fixing it at a reasonable value we assume that will not influence
a qualitative discussion on the phenomenon of SR, only the absolute values of
the
$\sigma$ correlation.
\end{itemize}

The $F(M,t)$ free energy of our system of $W$ coupled $S^z=\pm 1$ spins in
a $B(t)$ magnetic field can be computed using a mean-field approximation.
We assume that
\begin{equation}
F(M,t)=U(M)-T.S(M),
\end{equation}
with $U$ the internal energy, $T$ the temperature and $S$ the entropy. For
a cubic system, in the mean-field approximation we consider
\begin{equation}
U(M)=-3 J \frac{M^2}{W} + \mu_B M \cdot B(t),
\end{equation}
and:
\begin{equation}
S(M)=k \ln{( \left ( \begin {array}{c}
                      \frac{W+M}{2} \\ W
                      \end{array} \right )
            )}.
\end{equation}

The first and second order derivatives of $F(M,t)$ can be approximated
numerically. Equation (8) with the condition (10) can be now solved
also numerically for a given $C$ value. Starting from a $P(M,0)$ initial
distribution in principle we should be able to compute $P(M,t)$. The
$\sigma$ correlation can be calculated as
\begin{equation}
\sigma=\frac{1}{\tau} \int_0^{\tau} dt \int_{-W}^{W} M\cdot B(t)\cdot P(M,t) \:
dM,
\end{equation}
which is also computable.

Choosing the same units for $J$ and $B$ as in computer simulations we did
solved
numerically equation (8) with the normalization condition (10). $B(t)$ was
taken
in the form given by (2) and we fixed $J=20$. The "critical temperature" of the
system without magnetic field became in this way $T_c\approx 122$. We also
fixed
$W=40$ and considered different temperatures ($T$) around $T_c$, several
periods (P) and several amplitudes (A) for the oscillating magnetic field.
The constant $C\: dt$ was taken as $0.01$. We also checked that the overall
picture of $\sigma$ versus $T$ is not significantly influenced by lowering the
value of $C\: dt$.

Unfortunately the time scale of this theoretical approximation can not be
obviously related to the time scale used in computer simulations. Comparision
between theoretical and computer simulation results must be viewed under this
assumption.

\section{Results}

Our simulation results and theoretical approximations are summarized in Figs.
1-6.

In Fig.1 we present a characteristic computer simulation result for the shape
of the
$\sigma$ versus $T$ curve. One will observe that in accordance with the
predicted
phenomenon of SR, at a given $T_r$ temperature
$\sigma$ presents a maximum. The tail of this resonance peak is nicely
described
by a power law (bottom picture). Analysing the scaling exponents for different
simulations (different $P$, $A$ and $N$ values) we concluded that it lies in
the
[(-1.7)-(-2.4)] interval. A characteristic theoretical picture for the same
$\sigma$ versus $T$ curve is presented in Fig. 2. Again the resonance peak is
observed and
nice concordance with simulation results are suggested. Moreover the tail of
the
obtained resonance peaks are also described by power laws with scaling
exponents in the
[(-1.94)-(-2.05)] interval.

Figs. 3 and 4 presents the shape of the resonance peak for several values of
the
modulation amplitude $A$. Both theory (Fig. 4) and simulations (Fig. 3)
predicts
that the resonance temperature is almost independent of the modulation
amplitude,
exhibiting only very slight variation as a function of this (i.e. for higher
amplitudes $T_r$ is shifted in the direction of smaller values). In contrast
with
this, the height of the peak depends sensitively on the values of the
modulation
amplitude.

As we concluded also in [13], for small lattices the $T_r$ resonance
temperature is
strongly dependent on the lattice size (N), and in the limit of relatively big
lattices ($N\approx 20$) is tending to a constant limiting value. In this sense
for
three different modulation periods $P$, the simulation results are presented in
Fig. 5. We did not studied this dependence theoretically because our numerical
calculations were technically limited by the values of $W$ (i.e. only small
lattices
with $W<50$ were computable). From Fig. 5 we also learn that the temperature
$T_r$
is dependent on the modulation period $P$.

We studied the dependence of $T_r$ versus $P$ both by simulations and
theoretically.
The results in this sense are plotted in Fig. 6. Both simulations and theory
predicts that for high periods the $T_r$ resonance temperature is tending to
the
$T_c$ "critical temperature" of the system. This dependence can be described by
a
\begin{equation}
T_r-T_c=K_1 e^{-K_2 \: P}
\end{equation}
exponential law. Both for simulations and theoretical approximations the
step-like
form of the results are due to the fact that in detecting $T_r$ the temperature
was varied with steps of 5 units. The differences between the obtained $K_2$
values in
simulations and theory are due to the already mentioned problem that the time
scales
are not related each to the other.

\section{Conclusions}

The first conclusion would be that both our computer simulations and
theoretical data
suggest that the phenomenon of SR should be detected when one studies finite
ferromagnetic systems in oscillating magnetic fields. The characteristic peak
of
SR is obtained by studying the $\sigma=\mid<B(t)\cdot M(t)> \mid$ correlation
as
a function of temperature. For a given resonance temperature $T_r$, this
correlation
$\sigma$ exhibits a maximum.

Fixing the frequency, for small lattices ($W<4000$) the $T_r$ resonance
temperature
is dependent on the lattice sizes and is tending to a limiting value for
relatively
large ($W>4000$) lattices. The resonance temperature proved to be dependent
also on
the period of the magnetic field, and in the limit of large periods is
converging
exponentially to the critical temperature of the system. Because in real
experimental
conditions we are in the very high period limit (the time unit in simulations
is set by
$\tau$, the characteristic time for the flip of a spin), we expect $T_r$ to be
detected at $T_c$. We also concluded that the $T_r$ resonance temperature is
not
significantly influenced by the amplitude $A$ of the oscillating magnetic
field, the
value $A$ determining mainly the height of the resonance peak.

Our theoretical results proved to be in excellent concordance with the computer
simulations, justifying thus the considered approximations.

We consider that from the viewpoint of magnetism an experimental study of the
problem would also be of interest.

\section{Acknowledgment}

The work was done during a collaboration  with the University of Bergen
(Norway), supported by the Norvegian Research Council. I am grateful to
L.P. Csernai for his continuous help.

\newpage

\section*{References}
\begin{enumerate}
\item   R. Benzi, G. Parisi, A. Sutera and A. Vulpiani; {\em Tellus} {\bf 34},
	      10 (1982); SIAM {\em J. Appl. Math} {\bf 43}, 565 (1983).
\item    {\em Noise in Nonlinear Dynamical Systems}; eds. F. Moss and
	     P.V.E. McClintock (Campridge Univ. Press, Cambridge, England 1989)
\item    F. Moss; {\em Stochastic resonance: from the ice ages
	     to the monkey's ear} in: Some Problems in Statistical
		  Physics, ed. G. Weiss
		  (SIAM, Philadelphia 1992)
\item    P. Jung; {\em Phys. Rep.} {\bf 234}, 175 (1993)
\item    L. Gammaitoni, F. Marchesoni, E. Menichella-Saetta and
	       S. Santucci; {\em Phys. Rev. Lett.} {\bf 62}, 349 (1989)
\item    T. Zhou and F. Moss; {\em Phys. Rev. A} {\bf 41}, 4255 (1990)
\item    Proceedings on Nonlinear Dynamics in Optical Systems; eds.
	      M.B. Abraham, E.M. Garmire and P. Mandel (Optical Society of
		  America, Washington DC, 1991), Vol VII.
\item    L. Gammaitoni, M. Martinelli, L. Pardi and S. Santucci;
	      {\em Phys. Rev. Lett.} {\bf 67}, 1799 (1991)
\item    L. Gammaitoni, F. Marchesoni, M. Martinelli, L. Pardi
	       and S. Santucci; {\em Phys. Lett. A} {\bf 158}, 449 (1991);
\item   J. Heagy and W.L. Ditto; {\em J. Nonlin. Sci.} {\bf 1}, 423 (1991)
\item   P. Jung, U. Behn, E. Pantazelou and F. Moss; {\em Phys. Rev. A}
	{\bf 46}, R1709 (1992)
\item   A.R. Bulsara and G. Schmera; {\em Phys. Rev. E} {\bf 47}, 3734
	(1993)
\item   Z. N\'eda; {\em Phys. Rev. E}; {\bf 51} No. 6 (1995) (in press)
\item  M. Acharyya and B.K. Chakrabarti; in {\em Annual Reviews of
       Computational Physics I.},
	 edited by D. Stauffer (World Scientific, Singapore 1994)
\item   N. Metropolis, A.W. Rosenbluth, M.N. Rosenbluth, A.H. Teller and
	       E. Teller; {\em J. Chem. Phys.} {\bf 21}, 1087 (1953)

\end{enumerate}

\newpage
\begin{center}
\section*{Figure Captions}
\end{center}
\vspace{.30in}

$\:$ \\
\vspace{.15in}

{\bf Fig. 1.} Characteristic peak for the SR phenomenon obtained by
computer simulations. The bottom picture illustrates the
$\mid <B(t)\cdot M(t)> \mid=11.181 \cdot T^{-1.71}$ power law behaviour of the
tail ($m=M/W$, $N=20$, $A=40$, $P=100$ and $T_c=100$).

\vspace{.25in}

{\bf Fig. 2.} Characteristic shape of the SR peak given by our theoretical
approximation. The bottom picture illustrates the $\mid <B(t) \cdot m(t)> \mid=
2.088 \cdot T^{-1.96}$ power law behaviour of the tail ($m=M/W$, $W=40$,
$J=20$, $T_c=122$, $P=800$ and $A=1$).

\vspace{.25in}

{\bf Fig. 3.}  Computer simulation results for the shape of the SR peak
considering several values of the modulation amplitude $A$ ($m=M/W$, $N=20$,
$T_c=100$ and $P=50$).

\vspace{.25in}

{\bf Fig. 4.}  Theoretical results for the shape of the resonance peak
considering several values of the modulation amplitude $A$ ($m=M/W$, $W=40$,
$J=20$, $T_c=122$ and $P=800$).

\vspace{.25in}

{\bf Fig. 5.}  Computer simulation results for the dependence of the
resonance temperature ($T_r$) versus the lattice size ($N$). Results for three
different modulation periods ($P$) are presented ($T_c=100$ and $A=10$).

\vspace{.25in}

{\bf Fig. 6.}  Dependence of the resonance temperature [$\ln{(T_r-T_c)}$]
versus
the modulation period ($P$). The best fit line indicates the
$T_r-T_c=4.061 \cdot e^{-0.02\:P}$ and $T_r-T_c=4.312 \cdot e^{-0.003 \: P}$
dependence for simulation and theoretical data respectively (
for simulation: $A=10$, $N=20$; and for theory: $A=1$, $W=40$).

\end{document}